\documentclass[12pt,a4paper]{article}

\usepackage[margin=25mm]{geometry}
\usepackage[T1]{fontenc}
\usepackage[latin2]{inputenc}
\usepackage{ae}
\usepackage{graphicx}
\usepackage[american]{babel}
\usepackage{amsfonts}

\usepackage{amsthm}
\usepackage{amsmath}
\usepackage{amssymb}
\usepackage{wasysym}
\usepackage{paralist}
\usepackage{multirow}
\usepackage{cite}
\usepackage{pifont}
\usepackage{imakeidx}
\usepackage{arydshln}

\usepackage{subcaption}
\DeclareMathSymbol{\shortminus}{\mathbin}{AMSa}{"39}
\DeclareCaptionLabelFormat{subfig}{\figurename #1~\arabic{figure}/\alph{subfigure}:}
\makeatletter\def\p@subfigure{\thefigure/}\makeatother

\makeindex
\usepackage{color}
\definecolor{red}{rgb}{0.8,0,0}
\definecolor{green}{rgb}{0,0.6,0}

\usepackage{pgfplots}
\usepackage{hyperref}

\title{Application-oriented mathematical algorithms for group~testing}
\author{Endre Csóka\thanks{Supported by ERC Synergy grant No. 810115.}\\{\normalsize Alfréd Rényi Institute of Mathematics, Budapest, Hungary}}
\date{}

\begin{document}
\maketitle
\begin{abstract}

We have a large number of samples and we want to find the infected ones using as few number of tests as possible.
We can use group testing which tells about a small group of people whether at least one of them is infected.
Group testing is particularly efficient if the infection rate is low.
The goal of this article is to summarize and extend the mathematical knowledge about the most efficient group testing algorithms.
 focusing on real-life applications instead of pure mathematical motivations and approaches.


\end{abstract}

\tableofcontents

\newpage

\section{Introduction}

Group testing algorithms have a huge mathematical literature, but these focus on theoretically interesting idealized models.
This makes sense if we want to develop mathematical techniques, but these papers are too far from direct applications.
The current COVID-19 epidemic has created an urgent need for a more application-oriented summary and extension of our knowledge about testing algorithms, considering the mathematically ugly but practically important details.

\bigskip
\noindent
{\bf The mathematical model.}
We have $N$ number of {\bf samples}, each of them is {\bf positive} with a (small) probability $P$, independently, and the rest are {\bf negative}.
{\bf Pool} means a set of at most $K$ samples. We can a test a pool, and it outputs whether there is at least one positive sample in the pool.
The goal is to identify the positive samples with the least expected number of tests $T$. 
We can use an adaptive strategy, meaning that we can choose the pools depending on the earlier test results.

\bigskip
In this paper, we will use this mathematical model but also considering the following {\bf realistic refinements and goals}.

\begin{itemize}
\item We should not use the same sample for too many times and for too many {\bf rounds}. In more detail, the samples are continuously arriving at a speed of our choice (roughly), and we can make tests in rounds with a limited testing capacity (approx.\ 100) per round. We prefer getting faster results compared to the arrival time of the sample, and we also prefer using our full testing capacity.
\item There is a small probability of false test results, and the larger the pool size the lower the confidence. Moreover, the test results are not completely binary. Most test results are clearly positive or negative, but there are some likely positive or likely negative results with different confidence levels. Samples have a level of positivity, pooling takes approximately the average of these levels, and pools with lower positive levels are more likely to give negative (or more likely negative) test results.
\item We may be able to collect some preliminary information about the infection probabilities, and there are positive correlations between some samples, but collecting these information has a logistical cost.
\item The pooling algorithm will run partially manually, therefore we also try to offer algorithms which are easy to implement, and which are not too strongly dependent on the infection rate $P$ because we do not know $P$ exactly.
\end{itemize}

We think of $K$ as constant, we will use the default value $K = 10$, and $N$ as a large number, say, $N > 10$,000. (We will focus on the asymptotics $N \to \infty$, and we will apply approximations such as $\frac{\log_2 K}{\log_2 \log_2 K} \approx 2$.)

This model with no pool size limits (and without the realistic refinements) was first introduced by Ungár in 1960 \cite{ungar1960cutoff}.
In 1966, Katona used bounded-size pools for finding 1 infected person in a non-adaptive way \cite{katona1966separating}.
Since then there have been hundreds of papers published about similar questions.
However, for example, a recent paper by Malinovsky summarizes the open problems and how few we still know about the optimum even in this idealized model \cite{malinovsky2018conjectures}.
Also, the classical mathematical definition for the number of rounds is very different: it assumes that all samples arrive at the same time, we can make an unlimited number of tests per round, and we want to minimize the number of rounds for the entire algorithm.
These issues, and the lack of specification of the realistic parameters of the problem along with the rigorous mathematical approach were the reasons why no mathematical paper really aimed to introduce practically useful algorithms.

Very recently, a short medical paper was published about an appicable and applied solution for a non-adaptive case (without showing the exact algorithm) \cite{shental2020efficient}.
This also shows the urgent need of a bridge between combinatorics and application.
In contrast to their (unspecified) algorithm, our algorithms focus on a different trade-off by using much less number of tests but allowing positive tests to be identified in more than one rounds.

\subsection{Interpretation}

The quality of pooling is about reproducibility and not about reliability. It means that we are only speaking about positive and negative samples, and our analysis is independent from what the test result tells about the person being infected. Therefore, for us, {\bf having no false test results} (which we will not assume very strictly) only means the followings.
\begin{enumerate}
\item If we test the same single sample twice, then the two results should be the same. This (positive or negative) result is the type of the sample.
\item If we test a pool of size at most $K$, then the result is positive if and only if at least one of the samples is positive.
\end{enumerate}

For interpretation, a sample may refer to a joint sample of an entire household. Because combining their biological samples saves some testing resources, and if any person of a household gets a positive result, then the entire household should be put into quarantine anyway. This can be an important technique for application, and this affects the interpretation of the pool size limit $K$ and the infection rate $P$, but otherwise this is independent from the mathematical analysis of pooling.

``Next round'' in our model may mean the second next round in practice, because we may need some time for preparation of the pools. Calculating with 3 hours long Polymerase Chain Reaction (PCR) tests, one more round means 6 hours delay in the result.
Using two machines with starting times shifted by $1.5$ hours, one more round may mean only $4.5$ hours delay.


\section{A mathematical overview of efficient sample pooling algorithms depending on the infection rate} \label{sec:overview}

Finding the algorithm minimizing the expected number of tests in the strict mathematical sense can be extremely hard. (The nature of the problem is similar to the ``Subset Sum'' problem.) But it is easy to find nearly optimal algorithms. Therefore, out of them, we try to find the ones which are good at the other goals.

The information-theoretical lower bound tells that we always need at least
\begin{equation} \label{eq:entropy}
\big(- P \log_2(P) - (1-P) \log_2(1-P) \big) \cdot N \le T 
\end{equation}
number of tests. This bound is pretty sharp if $\frac{1}{K\log_2 K} < P < 60\%$.

\begin{figure}[!b]
\begin{center}
\begin{subfigure}[t]{0.47\textwidth}
\includegraphics[width=\textwidth]{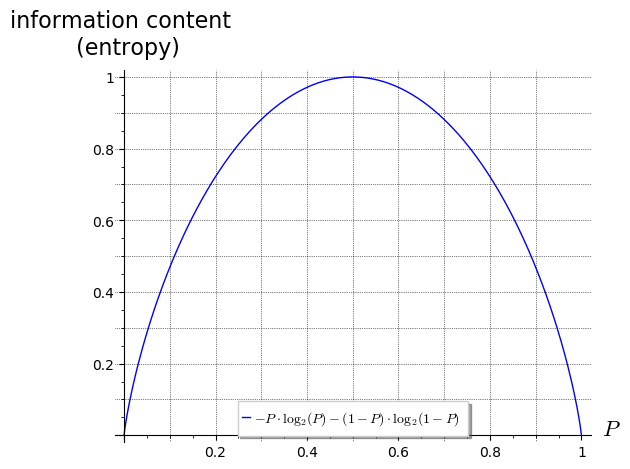}
\footnotesize{\caption{The information content of a binary variable (state of sample or test result) depending on the probability. E.g.\ if a test is positive with a probability $50\%$, then it gives 1 bit of information, and if a sample is positive with probability $10\%$, then it requires $0.47$ bit of information.} \label{fig:entropy}}
\end{subfigure}
\hskip 0.05\textwidth
\begin{subfigure}[t]{0.47\textwidth}
\includegraphics[width=\textwidth]{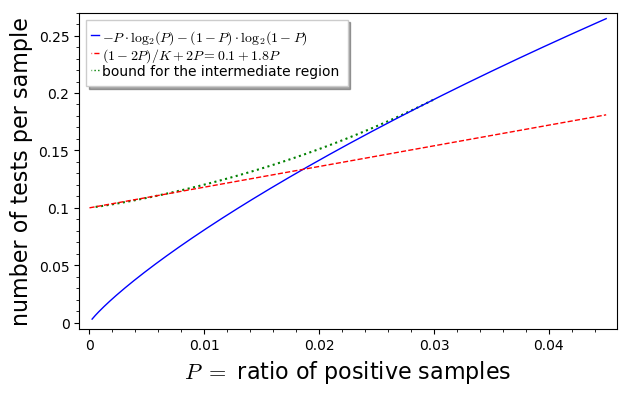}
\footnotesize{\caption{The figure shows the lower bounds for the tests per samples ratio depending on the rate $P$ of positive samples. The blue curve is the very same as on the left, and the red dashed line and green dotted curve assume $K = 10$.} \label{fig:bound}}
\end{subfigure}
\end{center}
\end{figure}

The proof of this bound rely on information theory, introduced by Shannon in 1948 \cite{shannon1948mathematical}. $- P \log_2(P) - (1-P) \log_2(1-P)$ is the so-called ``information content'' or entropy of the random type of one sample, depending on the probability $P$ of being positive. Assuming that the types are independent (which is not really true in practice), we need to get $N$ times this much information from the tests. A test can give at most 1 unit or 1 bit of information. It gives exactly 1 bit if and only if it has a probability 50\% for being positive (See Figure~\ref{fig:entropy}) given the earlier test results, and independently of the parallel tests.

But we cannot approximate this probability $50\%$ if $P$ is too low. For those cases, we need different bounds.

We have the following lower bound which is pretty sharp for $P < 1 / K^2$.

\begin{equation}
N \cdot \big( \frac{1-2P}{K} + 2P\big) \le T
\end{equation}

\noindent
(provided that $P \le 50\%$).
This bound is currently a conjecture, based on the fact that each negative sample needs at least one negative pooled test result, and most positive samples need at least two positive pooled test results. (Except with probability at most $1 / K$.)

For low but not very low probabilities, namely, for $\frac{1}{K^2} < P < \frac{1}{K\log_2 K}$, we have the following inequality: $2\frac{1-P}{K} \le \frac{T}{N} = t$ or
\begin{equation}
t \cdot \bigg( - \frac{1-P}{Kt} \log \frac{1-P}{Kt} - \Big( 1 - \frac{1-P}{Kt} \Big) \log \Big( 1 - \frac{1-P}{Kt} \Big) \bigg) \ge -P\log P - (1-P)\log(1-P).
\end{equation}
This bound is based on the fact that at least $\frac{(1-P)N}{K}$ test results must be negative, so if this is already more than half of the tests, then the probabilities cannot be close to 50\%.

\subsection{Summary and consequences} \label{sec:summary}

The maximum of these three lower bounds, as in Figure~\ref{fig:bound} is pretty sharp (for $0 < P < 60\%$), namely, we will consider algorithms using not much more than this number of tests. We can summarize these bounds and their consequences as follows.

If $P < \frac{1}{K\log K}$, then we should (almost) always use pools of size $K$, never use a sample in two parallel pooled tests, and it is somewhat better if we try to test pools with as close probabilities for being positive as possible. The lower $P$ is the more useful it is to have a larger pool bound $K$. In particular, if $P < \frac{1}{K^2}$, then we only need 1 test per $K$ samples plus approx.\ 2 tests per positive sample. For small $P$, having preliminary information about which samples are somewhat more likely positive is not very useful.

In contrast, for large $P$, the pool size limit $K$ is not important, because we can find an approximately optimal algorithm using smaller pool sizes. In this case, we can make use of the preliminary information about the individual probabilities of the samples, but it is enough to classify the samples into two or three classes, e.g.\ does or does not have any symptoms. We can make negligible use of any further refinement of the prior probabilities.

For a large infection rate $P$, constructing an almost efficient algorithm can be translated to the following simple goal. We should always use pools where the result has around 50\% chance of being positive, and the parallel tests should be almost independent. These are the only things we should care about in order to be efficient. But for practical use, I would suggest to aim for a better trade-off of about 40-50\%, because smaller probability implies smaller pools and using a sample in a smaller expected number a tests. The best trade-off depends on many parameters, but as long as these probabilities are always between 30\% and 60\%, the algorithm is still pretty good.

\section{Algorithms for a low infection rate $P \lessapprox \frac{1}{K^2}$}

\subsection{A very simple and safe algorithm} \label{sec:smallPsimple}

{\bf Number of tests}: less than $N/K + 3PN + K^2P^2 N < (1/K + 4P) N$.

\medskip
\noindent
A {\bf positive result} requires 2 rounds with a 3rd round for final confirmation.

\noindent
A {\bf negative result} requires typically 1, sometimes 2, very rarely 3 rounds.

\medskip
\noindent
\underline{Pros}: Easy to execute. It has a slightly higher tolerance for false test results, and gives a richer feedback about them.

\noindent
\underline{Cons}: It requires at least $P\cdot N$ more tests than the optimal one, and becomes much more inefficient with higher $P$.

\bigskip

\noindent
{\bf The algorithm.} In short, we test the samples in disjoint pools of size $K$, until a sample gets one negative or two positive pooled results. In the latter case, we make a confirmation test for that single sample.

In more detail, as soon as a sample has a negative pooled result, we report the sample as negative. If a sample has two pooled positive results, then we report it as likely positive, but we make a final confirmation test for that single sample, with no pooling. If a sample has one positive (and no negative) pooled result, then we call it a suspicious sample. In each round, beyond the confirmation tests, we choose the (fixed number of) disjoint pools of $K$ samples as follows. We distribute the suspicious samples into the pools as equally as possible. Then we complete each pool with different unexamined samples to a total of $K$ samples.

Note that if the pools contain more than 2 suspicious samples, then it signals a higher infection rate where this algorithm is less efficient.

In case if we get a positive pooled result, but later all $K$ samples get negative results, then it signals an error. Maybe the best way to fix it is to repeat the test which was positive, or putting back the $K$ samples to the list of unexamined samples.

\bigskip
\noindent
{\bf Improvement.} There are plenty of possibilities for little improvements, but the simplest one is the following. If a sample has two clear, strong positive results, and all of its fellows from the first positive pool got clear negative results in the second round, then we omit the confirmation test for that sample.

\subsection{An efficient algorithm}

{\bf Number of tests}: approx. $(1/K + 2P) N$ provided that $P < \frac{0.7}{K^2}$.

\medskip
\noindent
A {\bf positive result} requires typically 2, sometimes more rounds.

\noindent
A {\bf negative result} requires typically 1, sometimes more rounds.

\medskip
\noindent
\underline{Pros}: Very efficient with the number of tests.

\noindent
\underline{Cons}: Algorithmically more difficult. It is more sensitive for false test results.

\bigskip

\noindent
{\bf The algorithm.} Throughout the algorithm, the samples will be classified into 5 states: negative ($-$), unexamined (A), suspicios (B), likely positive (C), positive (+). States A, B, C express 0, 1, 2 active positive pool results, respectively. Initially, every sample is in state A. If a sample becomes likely positive (C), then we recommend putting them into quarantine immediately.

In every round, we choose the (fixed number of) pools by the following rule. We add one suspicious (B) sample into each pool until there is no more suspicious (B) sample. Or if suspicious samples arrive faster than this handling speed (which signals $P > 0.7 / K^2$), then we can allow 2 suspicious samples per pool (or we should change the strategy to the one in the next section).
We complete every pool with unexamined (A) samples to the size of $K$. If a test result is negative, then we change the state of all samples in it to negative ($-$). If a test result is positive, then each sample in it gets promoted to the next state: A $\to$ B or B $\to$ C. We keep the records of every positive test until either of the following event happens.
Out of the $K$ samples, if all but one samples $s$ are in the negative ($-$) state, then we change the state of $s$ to positive (+). Or if any of the samples are in the positive (+) state, then we relegate every other sample in it: C $\to$ B, B $\to$ A (we withdraw the promotion).

In case if we get a positive pooled result, but later all $K$ samples get negative results, then it signals an error. Maybe the best way to fix it is to repeat the test which was positive, or putting back the $K$ samples to the list of unexamined samples.

We may need to handle a very few samples which get stuck in state C. This is a minor issue except in the case of a higher infection rate P. We have multiple options as follows.

\medskip
\noindent
{\bf Version A.} (Simple.) We do nothing, namely, every person with likely positive (C) sample stays in quarantine.

\medskip
\noindent
{\bf Version B.} (Safe.) If a sample stays in state C for a fixed number of rounds, or at the end, we make a confirmation test for that single sample.

\medskip
\noindent
{\bf Version C.} (Efficient.) If we see a deadlock, which is a cycle in the hypergraph of likely positive (C) samples and positive pools, then we test every second or third sample in the cycle. (Even more efficient if we test them in a pool with $K-1$ unexamined (A) samples.)

\medskip
\noindent
{\bf Version D.} An adaptive refined combination of the previous strategies, especially if the logistics is directed by computer.

\section{Efficient algorithms for infection rate $\frac{1}{K^2} < P < \frac{1}{2K}$}

\subsection{Efficient algorithms for infection rate $\frac{1}{K^2} < P < \frac{1}{5K}$}

{\bf Number of tests}: approx. $(1/K + 4P) N$, but the factor of 4 depends on $P$ and the version of the algorithm.

\medskip
\noindent
A {\bf positive result} requires approx. 4 rounds.

\noindent
A {\bf negative result} requires 1 - 2 rounds in average.

\medskip
\noindent
\underline{Pros}: Very efficient with the number of tests.

\noindent
\underline{Cons}: Algorithmically difficult.

\bigskip

\noindent
{\bf The algorithm.} The samples will be classified into 6 states: negative ($-$), unexamined (A), suspicios (B), very suspicious (C), likely positive (D), positive (+). States A, B, C, D express 0, 1, 2, 3 active positive pool results, respectively. Initially, every sample is in state A. If a sample becomes very suspicious (C), or at latest when it becomes likely positive (D), then we recommend putting them into quarantine immediately.

In every round, we choose the pools by the following rule. We make pools using all suspicious (B) and very suspicious (C) samples, and with as many unexamined (A) samples as needed to make pools of size $K$, and we make it with as equal probabilities for the pool being positive as we can.

In particular, in the case of $P < \frac{1}{2K\sqrt{K}}$, for each very suspicious (C) sample, we make a pool with this sample and $K-1$ unexamined (A) samples. For the rest of the pools, we add all suspicious (B) samples into the pools as evenly as possible (preferably without adding any two samples to the same pool if these samples are from the same positive pool). We complete the pools with unexamined (A) samples to size of $K$. If a test result is negative, then we change the state of all samples in it to negative ($-$).

 If a test result is positive, then each sample in it gets promoted to the next state: A $\to$ B, B $\to$ C or C $\to$ D. We keep the records of every positive test until either of the following event happens. If all but one sample $s$ are in the negative ($-$) state, then we change the state of $s$ to positive (+). Or if any of the samples are in the positive (+) state, then we relegate every other sample in it: D $\to$ C, C $\to B$ or $B \to A$ (we withdraw the promotion).

We may need to handle a very few samples which get stuck in state D. This is a minor issue except in the case of a higher infection rate. We have multiple options as follows.

\medskip
\noindent
{\bf Version A.} We do nothing, namely, every person with likely positive (D) sample stays in quarantine.

\medskip
\noindent
{\bf Version B.} If a sample stays in state D for a fixed number of rounds, or at the end, we make a confirmation test for that single sample.
The larger $P$ the less we should wait with confirmation.

\medskip
Or somewhat independently of these versions, we can use one of the following variants of the algorithm.

\medskip
\noindent
{\bf Variant 1.} A safer but less efficient variant is that we change the status of a sample $s$ to positive (+) only if not one but two records of positive pools imply that $s$ must be the positive sample.

\medskip
\noindent
{\bf Variant 2.} If the pooling algorithm is directed by a computer, then these algorithms (including the one in the previous section) can be very slightly improved, and classes A, B, C, D could be replaced essentially with probabilities. In each round, we should make pools of approximately equal probabilities of being positive (any reasonable heuristical algorithm would be good here). This could also handle the case of non-binary test results but different levels of ambiguity in them.

\subsection{Extension to the case $\frac{1}{5K} < P < \frac{1}{2K}$}

In this parameter regime, the best algorithm depends on many factors, and we have to choose between efficiency and simplicity.
But in the case if the pool size limit $K$ is critical, and smaller pools would provide better confidence, then probably the best solution is decreasing $K$ from the 2nd or 3rd round, hereby testing smaller pools.
In addition or alternatively, for the samples reaching the ``likely positive'' (D) state, we consider them as an input of an algorithm designed for a higher infection rate. (This is a generalization of Version B.)

But if computer support is available, then we can slightly improve the algorithm according to Variant 2, or following the guidelines in Section~\ref{sec:summary}.

\section{Simple and universal algorithms} \label{sec:simple}

\subsection{A simple, fast and safe but less efficient algorithm}

{\bf Number of tests}: approx.\ $2\sqrt{P} \cdot N$ if $P \gtrapprox \frac{1}{K^2}$, otherwise $N \cdot \big( 1 + \frac{1}{K} - (1-P)^K \big)$.

\medskip
\noindent
A {\bf positive result} requires 2 rounds.

\noindent
A {\bf negative result} requires 1 or 2 rounds.

\medskip
\noindent
\underline{Pros}: Better tolerance for test errors, easy to execute, fast result, universal with no assumption on the infection rate.

\noindent
\underline{Cons}: Not very efficient with the number of tests, especially around the infection rate $P \approx \frac{1}{K\log_2(K)}$. It may use twice (more than $\sqrt{\frac{K}{\log_2(K)}}$ times) as much tests as the most efficient algorithm.

\bigskip

\noindent
{\bf The algorithm.} 
Let $k \approx \min\Big(K,\ \sqrt{\frac{1}{P}}\Big)$.
Test the new samples $k$ by $k$ in disjoint pools, and if a result is positive, then in the next round, test all samples in the pool one by one.
(If none of these $k$ tests are positive, then it signals a test error. In this case, we could simply repeat the procedure from testing again the same pool of $k$ samples.)

\medskip
\noindent
{\bf Version 1.}
For a positive pool of $k$ samples, we test only $k-1$ out of the $k$ samples in the next round.
If any of them is positive, then the $k$th sample should be put back to the list of unexamined samples (and maybe make sure that next time it will not be the $k$th sample again).
If all of them are negative, then the $k$th one must be positive. Optionally, we can make a confirmation test next round, but we should not wait for it with reporting the positive result.

\medskip
\noindent
{\bf Version 2.} After a positive pooled result of $K$ samples, in the next round we test them in disjoint pools of size $L = \lceil \sqrt{K} \rceil$ by $L$, leaving approximately $L$ samples untested. Then in the third round, we test the individual samples which got the second positive pooled result. 
About the samples which were untested in the second round, we test them in the third round individually if none of the pools in the second round were positive, otherwise we put them back to the list of unexamined samples.

\medskip
\noindent
{\bf Version 3.} We can modify Version 2 so that we test some pools in the second round so that some samples are tested in multiple pools. This can provide more efficient algorithms, but the algorithm depends on the level of disadvantages of using larger pools, on the rate of false test results, and on $K$ and $P$.

\subsection{An improvement of a classical algorithm} \label{sec:classic}

This algorithm is primarily suggested for those who are familiar with the well-known but {\color{red} \mbox{INEFFICIENT} ALGORITHM SHOWN IN THE FIGURE}.
Our algorithm is an improvement of that one.

\begin{figure}[!b]
\begin{center}
\includegraphics[width=\textwidth]{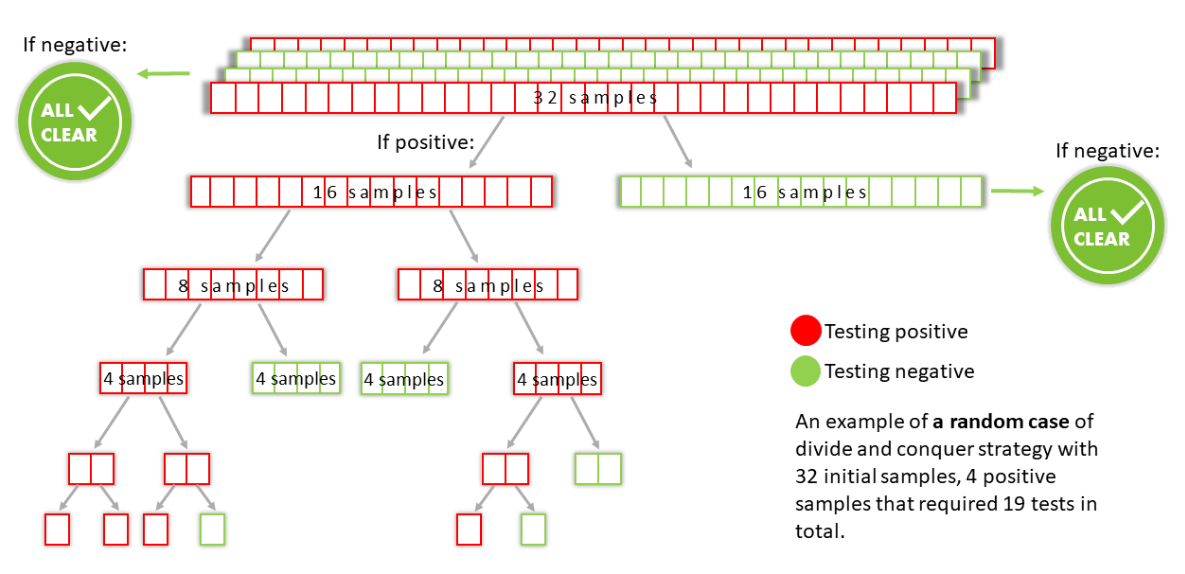}
\caption{{\color{red} DO NOT USE THIS ALGORITHM!} This is a visualization of the advantages of pooling, and it shows a well-known but inefficient algorithm. Our algorithm in Section~\ref{sec:classic} is an improvement of this one.
(For example, always dividing into not 2 but 3 parts would already improve this algorithm.)
The figure is taken from \href{https://medium.com/@dinber19/more-with-less-using-pooling-to-detect-coronavirus-with-fewer-tests-8ba1a2cd8b67}{medium.com}.}
\end{center}
\end{figure}

\bigskip
\noindent
{\bf Number of tests}: close to the entropy bound \eqref{eq:entropy} if $P \gtrapprox \frac{1}{2K}$, otherwise at most $1/K + P\log(K)$.

\medskip
\noindent
A {\bf positive result} requires $\log_2(k) + 1$ rounds in average, worst case is $2\log_2(k)$ rounds.

\noindent
A {\bf negative result} requires 2 rounds in average, but the worst case is $2\log_2(k)$ rounds.

\medskip
\noindent
\underline{Pros}: Universal with no assumption on the infection rate, very efficient if $P > \frac{1}{2K}$. Easy to execute for those who are familiar with the classical inefficient algorithm.

\noindent
\underline{Cons}: Requires too many rounds. It is not efficient with the number of tests for infection rate $P < \frac{1}{2K}$. Around the infection rate $P \approx \frac{1}{K\log_2(K)}$, it may use up to 20\% more tests as the most efficient algorithm.

\bigskip

\noindent
{\bf The algorithm.} For $k \approx \min \big( K,\ \log(0.55) / \log(1-P) \big)$ (where ``$0.55$'' is $1 - ($the approx.\ 40-50\% mentioned in Section~\ref{sec:summary})), we test the samples $k$ by $k$ in disjoint pools, and if a result is positive, then we repeat the following procedure starting with these $k$ suspicious samples in $X$. As long as we have more than 1 suspicious samples in $X$, we test a pool with half of them, rounded down (preferably choosing the samples which arrived earlier). If positive, then we continue the procedure with this half of samples $X_1$ and the other half $X_2 = X \setminus X_1$ should be put back to the list of unexamined samples. If negative, then we continue the algorithm with the other half of samples $X_2$. If 1 suspicious sample remains in $X$, then we report it positive.

\medskip
\noindent
{\bf Variant 1.} If a test result was not very clear, or we continued with the second half of samples for too many consequtive rounds, or by any reason we start to have doubt that there is indeed a positive sample in the current set of suspicious samples, then next round we test a pool with more than half of the suspicious samples, but at most all but one of them. If the test result of all but one suspicious samples was still negative, then next round we make a confirmation test for that only remaining sample.

\medskip
\noindent
{\bf Variant 2.} Independently from Variant 1, if we see that a halved pool $X_1$ was positive, but less strongly positive than the bigger pool $X = X_1 \cup X_2$, then test the other half $X_2$, as well. This variant is not recommended for $P < \frac{1}{2K}$.

\section{More efficient algorithms for higher infection rates}

The best algorithm is highly dependent on the fine parameters of the problem and our preferences, as listed in the introduction.
We list the most likely suggestions considering efficiency, simplicity and fast result, focusing on the case if we have no preliminary information about which samples are more likely positive. We list them with decreasing order of $P$. Section~\ref{sec:overview} helps finding a pretty efficient algorithm considering any possible parameters and preferences.

\bigskip
\noindent
{\bf Phase 1.} If $P > 38\%$ (or in the idealized model, $P > \frac{3-\sqrt{5}}{2} \approx 38.2\%$) then we should test the samples one by one. This is what we call the trivial algorithm.
(If there are disadvantages of pooling, then it decreases the threshold, while positive correlations slightly increase it. According to my knowledge, this is the only threshold which is mathematically proved, by Ungár in 1960.)

\bigskip
\noindent
{\bf Phase 2.} For $23\% \le P \le 38\%$, we should use pools of size 2. (We use the same algorithm as shown in Section~\ref{sec:simple} for $k = 2$.)
If the result is positive, then the safer and faster but less efficient option is testing both samples in the next round.
But, especially for lower $P$ and less correlation between two neighboring samples, we can improve it by the following modification.

If we pool the samples $(1, 2)$, and result is positive, then in the next round, we test only sample $1$. If the result is negative, then we know that sample $2$ must be positive. Or if sample $1$ is positive, then we test the pool $(2, 3)$ with a new sample $3$, because sample $2$ is not more suspicious than an unexamined sample. This algorithm saves 11.2\% of the tests at an infection rate $P = 30\%$ and $18.9\%$ at rate $P = 25\%$, in the idealized model.
(We conjecture that this algorithm minimizes the number of tests for $27\% < P < 38\%$.)

\medskip
\noindent
{\bf As a more generaly tool}, if we know that at least one of samples 1 and 2 must be positive, and this and the prior probabilities are all we know about them, then in the next round, we test only sample $1$. If the result is negative, then we know that sample $2$ must be positive. Or if sample $1$ is positive, then we put back sample $2$ to the list of unexamined samples.

\bigskip
\noindent
{\bf Phase 3.} For $15\% \le P \le 24\%$, we suggest the following algorithm. We start with pools of size 3. If a result is positive with samples $(1, 2, 3)$, then in the next round, we take an unexamined sample $4$ and we test the pools $(1, 4)$ and $(2, 4)$.

If both results are negative, then we report that sample $3$ is positive, samples 1, 2, 4 are negative.

If, say $(1, 4)$ was positive but $(2, 4)$ was negative, then we report that sample $1$ is positive, samples $2$ and $4$ are negative, and we put sample $3$ back to the list of unexamined samples.

If both are positive, then we test samples $1$ and $2$ alone. If both are positive, then we put samples $3$ and $4$ back to the unexamined tests. If one is positive and the other is negative, then we report sample $4$ to be positive and we put sample $3$ back to the unexamined tests. If both are negative, then we report both samples $3$ and $4$ as positive.

At an infection rate of 20\%, this algorithm saves $27.6\%$ of the tests compared to the trivial algorithm, and $4.2\%$ compared to the algorithm in Phase 2.

\bigskip
\noindent
{\bf Phases 4-5.} For $10\% < P < 17\%$, we test the samples 4 by 4 (for $14\% < P < 17\%$) or 5 by 5 (for $10\% < P < 14\%$), and if positive, then we test a pool of 2 of these samples. If also positive, then we finish in the same way as in Phase 2, and the other 2 or 3 samples should be put back to the list of new samples. If negative, then we finish with the remaining 2 or 3 samples as in Phase 2 or 3, respectively.

\bigskip
\noindent
{\bf Phases 7-8.} For $7\% < P < 10\%$, we test the samples 7 by 7 (for $8\% < P < 10\%$) or 8 by 8 (for $7\% < P < 9\%$), and if positive, then we test two pools of 3 samples: $(1, 2, 3)$ and $(1, 4, 5)$. If both are negative, then we use the algorithm in Phase 2 or 3 for the other 2 or 3 samples. If one of them is positive, then finish with the algorithm in Phase 2 with the two suspicious samples.
If both $(1, 2, 3)$ and $(1, 4, 5)$ are positive, then we test the pool $(1, 2)$, and if negative, then sample 3 must be positive and the others should put back to the list of unexamined samples, or if $(2, 3)$ is positive, then we apply the algorithm in Phase~2 for samples 2 and 3, and Phase~3 for samples $1, 4, 5$.

(Optionally: if both $(1, 2, 3)$ and $(1, 4, 5)$ are positive and $(1, 4, 5)$ was the more strongly positive one, then we test the pool $(4, 5)$ instead.)

\bigskip
\noindent
{\bf Phase 10.} For $5\% < P < 7\%$, we test the samples 10 by 10 (for lower $P$), and if positive, then we test two pools of 4 samples: $(1, 2, 3, 4)$ and $(1, 5, 6, 7)$. If one or both of them are negative, then we get 3 suspicious samples, and we use the algorithm in Phase 3 for them. If both $(1, 2, 3, 4)$ and $(1, 5, 6, 7)$ are positive, then we test the pool $(2, 3, 4)$. If negative, then sample 1 must be positive and samples $5, 6, 7$ should put back to the list of unexamined samples. Or if $(2, 3, 4)$ is positive, then we apply the algorithm in Phase~3 for samples $2, 3, 4$, and Phase~3 for samples $1, 5, 6, 7$.

(Optionally: if both $(1, 2, 3, 4)$ and $(1, 5, 6, 7)$ are positive and $(1, 5, 6, 7)$ was the more strongly positive one, then we test the pool $(5, 6, 7)$ instead.)

\section*{Acknowledgements}

I want to say thank you to Viktória Lázár, Péter Vilmos and the \href{http://koronavirusszures.com}{Suppress19 group} for practical information about pooled testing.

\bibliography{pooling}

\end{document}